# A Sharp Event in the Image A Light Curve of the Double Quasar 0957+561 and Prediction of the 1996 Image B Light Curve

Tomislav Kundić, Wesley N. Colley [1], J. Richard Gott, III, Sangeeta Malhotra, Ue-Li Pen, James E. Rhoads, Krzysztof Z. Stanek, Edwin L. Turner

Princeton University Observatory, Peyton Hall, Princeton, NJ 08544

lens@astro.princeton.edu

and

Joachim Wambsganss

Astrophysikalisches Institut Potsdam, An der Sternwarte 16, 14482 Potsdam, Germany

## ABSTRACT

CCD photometry of the gravitational lens system 0957+561A,B in the $g$ and $r$ bands was obtained on alternate nights, weather permitting, from December 1994 through May 1995 using the Double Imaging Spectrograph (DIS) on the Apache Point Observatory (APO) 3.5-meter telescope. The remote observing and fast instrument change capabilities of this facility allowed accumulation of light curves sampled frequently and consistently. The Honeycutt ensemble photometry algorithm was applied to the data set and yielded typical relative photometric errors of approximately 0.01 magnitudes. Image A exhibited a sharp drop of about 0.1 magnitudes in late December 1994; no other strong features were recorded in either image. This event displays none of the expected generic features of a microlensing-induced flux variation and is likely to be intrinsic to the quasar; if so, it should also be seen in the B image with the lensing differential time delay. We give the expected 1996 image B light curves based on two values of the time delay and brightness ratio which have been proposed and debated in the literature. Continued monitoring of the system in the first half of 1996 should easily detect the image B event and thus resolve the time-delay controversy.

*Subject headings:* quasars: 0957+561, cosmology: gravitational lensing

---

[1]Supported by the Fannie and John Hertz Foundation

## 1. Introduction

The double quasar 0957+561A, B was the first gravitational lens to be discovered (Walsh, Carswell & Weymann 1979). Its redshift is $z_Q = 1.41$, the separation of the two images is $6.1''$ and the lensing galaxy is at the center of a $z_L = 0.36$ cluster of galaxies. Shortly after the discovery, changes in brightness of the two quasar images were detected and measured by several groups (Lloyd 1981; Miller, Antonucci & Keel 1981; Keel 1982; Schild & Weekes 1984; Schild & Cholfin 1986). The double quasar 0957+561A, B was also the first system for which an attempt to measure the time delay was reported [see, e.g., Dyer & Roeder (1980) or Florentin-Nielsen (1984)]. Since the gravitational time delay is inversely proportional to the Hubble parameter $H_0$ (Refsdal, 1964), it can be in principle used to determine $H_0$. For the system 0957+561 with its many observational constraints (separation of images, intensity ratios, radio structure on VLA scale, VLBI jets, X-ray properties) a variety of lens models exist (Young et al. 1981; Narasimha, Subramanian, & Chitre 1984; Falco, Gorenstein & Shapiro 1991; Kochanek 1991).

There is, however, a degeneracy between smoothly distributed matter of the underlying galaxy cluster and the matter in the lensing galaxy itself (Gorenstein, Falco & Shapiro 1988). Further uncertainty in the determination of $H_0$ in the double quasar arises as a result of a controversy in the measured value of the time delay between different groups. While optical data typically yielded a time delay of $\sim 1.1$ years, radio observations revealed a longer time delay of $\sim 1.5$ years (Borgeest & Refsdal 1984; Falco, Gorenstein, & Shapiro 1985, 1991; Vanderriest et al. 1989; Schild 1990; Kochanek 1991; Rhee 1991; Lehár et al. 1992, Press, Rybicki & Hewitt 1992b; Schild & Thomson 1995; Pelt et al. 1994, 1995). Prompted by this disagreement, we decided to include 0957+561 into our optical monitoring program of gravitational lenses and gravitational lens candidates.

The double quasar (among other lenses) has been monitored almost continuously from the beginning of December 1994 through the end of May 1995 with the 3.5m Apache Point Observatory (APO) telescope. Weather and equipment permitting, we obtained two to four images of 0957+561 in $g$ and $r$ bands every other night. The acquisition and reduction of the CCD data is discussed in §2. In §3 we present the observed light curves and a smooth fit through the data using the method of Press et al. (1992a). We also plot the predicted image B light curve for 1996 derived from the observed image A light curve and the two widely accepted values of the time delay. In §4 we contrast the intrinsic quasar variability with the microlensing-induced features. Finally, in §5 we give a brief discussion of our results and their implications.

## 2. Data Acquisition and Reduction

Images were obtained at the APO[2] 3.5-meter telescope with the Double Imaging Spectrograph (DIS) in the imaging mode.

The Double Imaging Spectrograph has two independent collimators and cameras for the blue and red side. The incoming light is split by a dichroic with a transition wavelength of 5350 Å. On the blue side the detector is a thinned, uv-coated SITe (formerly Tektronics) 512x512 CCD, and on the red side it is a thinned 800x800 TI chip. The measured scales are 1.086 ″/pixel in the blue and 0.610 ″/pixel in the red (Lupton, 1995). Gunn-Thuan $g$ and $r$ filters were used on the blue and red sides for the lens monitoring program. Since the filter wheel blocks some of the light, the effective area on the red chip was reduced to $384'' \times 256''$ and on the blue chip to $396'' \times 266''$.

The gain and linearity of the chips were recently measured by Gloria (1995). On the blue side the gain is 1.00 electrons/ADU; on the red side it is 1.83 electrons/ADU. The blue chip is linear to 61000 ADU and the red chip is linear to 48000 ADU. These figures were used to exclude overexposed stars on the $g$ and $r$ frames, respectively.

Lens monitoring was scheduled every other night from December 2, 1994 to May 31, 1995. All data was acquired by operating the telescope remotely from an observing station in Princeton. Several images with different exposure times were taken to encompass comparison stars with a wide magnitude range. The telescope was offset by 10–20″ between exposures to minimize spurious variations from bad pixels and to allow construction of sky flats from a median of program exposures. Images were taken mostly in non-photometric conditions with typical seeing FWHM between 1.3″ and 1.8″. Roughly 20 to 45 minutes were devoted to lens monitoring each night, of which perhaps one third was used to obtain the 0957+561 data; if not already in use, DIS was mounted just for these observations.

CCD images were processed using IRAF's *ccdred* package. A "superbias" was created from a median of all bias images taken throughout the season and subtracted from each program exposure. The remaining variations in the mean bias level were subtracted by fitting a low-order spline to the overscan columns. Pixel-to-pixel sensitivity variations were removed using a sky flat field obtained by medianing two weeks of monitoring data. Images were grouped according to the moonphase, with boundaries at first and third quarter. At least 14 exposures were combined for each sky flat.

---

After the preliminary data processing, aperture photometry was performed on the two quasar images and 9 comparison stars. The aperture radius was limited to $3''$ by the separation of the quasar images. A modal value of the sky from an annulus around each object was subtracted from the aperture flux. Objects with the peak flux exceeding the linearity threshold were flagged at this stage, as well as those with S/N below 50. Images with 3 or more comparison stars were retained for differential photometry, resulting in 183 good $r$ exposures and 167 good $g$ exposures.

Light curves for the two quasar images in $g$ and $r$ bands were derived using the inhomogeneous CCD ensemble photometry algorithm developed by Honeycutt (1992). Beyond selection criteria described above, we did not require a fixed set of comparison stars for "strict" ensemble photometry.

Following the notation of Honeycutt (1992), we briefly describe the least square solution method for measuring instrumental magnitudes of $ss$ stars in a set of $ee$ exposures. The instrumental magnitude of star $s$ on exposure $e$ is given by

$$m(e, s) = m0(s) + em(e) \quad , \qquad (1)$$

where $m0(s)$ is the mean instrumental magnitude of star $s$, and $em(e)$ is the exposure magnitude for exposure $e$. The exposure magnitude represents exposure-to-exposure variations common to all stars due to exposure time, atmospheric extinction, etc.

The quantity to be minimized is then

$$\beta = \sum_{e=1}^{ee} \sum_{s=1}^{ss} \left[ m(e, s) - m0(s) - em(e) \right]^2 w(e, s) \quad , \qquad (2)$$

where $w(e, s)$ is the weight given to instrumental flux of star $s$ in exposure $e$. This weight is set to 0 if star $s$ is overexposed or missing. Otherwise, each well-exposed star is initially given the same weight $w(e, s) = 1$, which is later refined based on exposure-to-exposure variance of the instrumental flux of star $s$ ($\sigma[m0(s)]$) and star-to-star instrumental magnitude variation in exposure $e$ ($\sigma[em(e)]$). Note that quasar images A and B are not included in the least square sum $\beta$ because of their expected intrinsic variation.

Plots of $\sigma[m0(s)]$ versus $m0(s)$ and $\sigma[em(e)]$ versus $em(e)$ can be also used to detect variable objects and bad exposures, respectively. In Fig. 1 we plot $\sigma[m0(s)]$ versus $m0(s)$ for $g$ and $r$ instrumental magnitudes. Clearly, the A and B quasar images are more variable than the comparison stars with similar magnitude. The variances of instrumental $g$ and $r$ magnitudes of quasar images A and B plotted in Fig. 1 include both photometric errors and intrinsic quasar variability. A variance of the quasar magnitude that represents photometric errors only is obtained by fitting a functional form to the $\sigma[m0(s)]$ vs. $m0(s)$ relation of comparison stars and evaluating it at the quasar magnitude. An implicit assumption in this procedure is that the quasar magnitudes are measured with the same precision as the comparison star magnitudes. This results in a somewhat optimistic error estimate, because photometry of the quasar images is complicated by their proximity (overlapping PSFs) and the seeing-dependent light contribution of the galaxy underlying image B. These sources of errors have been thoroughly discussed by Schild & Cholfin (1986), who conclude that each effect contributes no more than 1% error in seeing better than $2''$.

Approximate absolute magnitude offsets in the two bands were calculated by comparison of instrumental $g$ and $r$ magnitudes with the data on 5 stars in the 0957+561 field published in Table 1 of Schild & Cholfin (1986). The rms scatter in relative magnitudes between the two datasets was 0.010 mag in $r$ and 0.017 mag in $g$.

Since we have data in two bands, we could in principle include a color term in the Eq. 1. The exposure magnitude of star $e$ with mean instrumental $g$ and $r$ magnitudes $m0_g(s)$ and $m0_r(s)$ would then be $m_r(e, s) = m0_r(s) + em_r(e) + ec_r(e)[m0_r(s) - m0_g(s)]$, where $ec_r(e)$ is the exposure color of exposure $e$. In practice, the additional term $ec_r(e)$ did not reduce the sum of squares significantly and was removed from further consideration.

## 3. Resulting Light Curves and Predictions

Light curves for the $g$ and $r$ images of the QSO 0957+561 A, B are shown in Fig. 2. These light curves are characterized by high temporal resolution (data taken every other night when clear) and good photometric precision (0.01 mag for bright comparison stars, 0.02 mag for faint comparison stars and quasar images). Observations from the same night were combined to a single point on the light curve using a weighted average of individual exposures. The solid line fit to the data is described below.

In the first month of observing (December 1994) image A exhibited a sharp chromatic drop in instrumental flux over a period of about two weeks. The approximate amplitude of this event was 0.08 mag in $r$ and 0.13 mag in $g$. Because of the chromaticity and short duration of the event, we believe it represents an intrinsic variation of quasar flux that will repeat in image B after a 1.1-1.5 year time delay. Careful monitoring of both images in the first half of 1996 could then be used to accurately measure the gravitational time delay $\tau$ in this lens system and settle the long-standing controversy about the exact value of $\tau$.

Even after careful data reduction, light curves in Fig. 2 show correlated variations in instrumental magnitudes of images A and B with an amplitude of 0.01-0.02 mag. A possible reason for such correlations is time variation of the residual large-scale structure in the sky flats used to remove pixel sensitivity variations across the field. Since quasar images are close to each other and preferentially located near the center of each frame, they will be flat-fielded differently from the comparison stars scattered across the frame. Another culprit might be seeing-induced variations in instrumental flux caused by overlapping PSFs of the two quasar images and the underlying galaxy. In any case, these systematic photometric errors are comparable to the statistical ones and small compared to the December 1994 image A event.

In order to clearly separate spurious correlated variations of images A and B from the true variability of the quasar, we plot the difference in instrumental $r$ and $g$ magnitudes of the two images in Fig. 3. It is clear from this plot that the December event occurred only in image A, even though its amplitude relative to the image B flux is reduced to about 0.05 mag in $r$ and 0.10 mag in $g$.

We fit a continuous curve to the data following Press, Rybicki & Hewitt (1992a), who suggest an "optimal reconstruction of irregularly sampled data." Their method estimates a covariance function of magnitude as a function of temporal separation in the observational sampling, which allows a fit that globally minimizes $\chi^2$. We follow their method rather strictly, except in our assumed covariance function. Instead of the single power-law they suggest, we used a power-law softened at high frequency to reduce strong spikes or kinks caused by any single outlier in the data. We have taken a power-law covariance function of the form $V(|t_1 - t_2|) \propto (|t_1 - t_2|/\text{day})^{0.4}$, smoothly softened to a power-law with index 0.2 over the range of 5 days $< |t_1 - t_2| < 15$ days.

The solid line in Fig. 2 displays our fits for the $r$ and $g$ monitoring data. In each band, there is a significant drop in brightness of image A, with no corresponding drop in image B. In $g$ the drop of about 0.13 magnitudes should be recognizable if it repeats in B. We thus predict in Fig. 4 the future shape of the $g$ light curve in B for two different time-delays, 536 days, obtained by analysis of radio data (Lehár et al. 1992) and a combined optical-radio sample (Press et al. 1992b), and 415 days obtained by Vanderriest et al. (1989) and Schild & Thomson (1995). The latter time delay is also advocated by Pelt et al. (1994, 1995).

Notice also the offset in the ordinate for the two curves in Fig. 4. Just as different methods of analysis yield different time delays $\tau$, they also yield different magnitude offsets between $A(t)$ and $B(t - \tau)$, (i.e., the intrinsic difference in brightness between the two images). Press et al. (1992b) suggest an offset of 0.0950 magnitudes for the optical data, while Vanderriest et al. (1989) suggest one of 0.03 magnitudes. The discrepancy is easily visible in the plot, and should serve as a further diagnostic for distinguishing the two time delay estimates at the time of event in image B. We have tacitly assumed that the image flux ratios are constant from one optical band to another, neglecting reddening in the lensing galaxy and the light contribution of this galaxy to the measured instrumental fluxes.

## 4.  The Microlensing Hypothesis

There can be two causes for the observed variability of the quasar image: it can reflect an internal change in the luminosity of the quasar or it can be a result of gravitational microlensing by compact objects along the line of sight to the quasar. Only intrinsic variability is expected to repeat in image B and can thus be used to measure the gravitational time delay. While it is possible that the feature is microlensing-induced, we argue that this scenario is unlikely because of chromaticity, timescale and shape of the event. These arguments are considered in turn.

An important indication that the observed event is caused by intrinsic quasar variability is its chromaticity. The event amplitude in $g$ (0.13 mag) is almost twice as large as in $r$ (0.08 mag), in contrast to the expected achromaticity of gravitational lensing. There is a possibility of measuring slightly different light curves in two bands if inner parts of the quasar accretion disk are "hotter" or "bluer" than the outer parts (Wambsganss & Paczyński 1991), but this second order effect is not expected to produce differences as big as seen here. In fact, for effective source size ratios of 1.26 or 1.59 between the continuum in $g$ and $r$, Wambsganss & Paczyński (1991) find that the expected change in color for a "smooth" event is less than 10% of the change in $g$, much smaller than what we observed.

The short duration of the event also disfavors the microlensing hypothesis. The typical microlensing time scale in the system 0957+561 for solar-mass stars is expected to be of order 100 years (Young 1981). For much less massive objects, Gott (1981) estimated the expected microlensing time scale $\tau_{\mathrm{ML}}$ in 0957+561 to be roughly

$$\tau_{\mathrm{ML}} \approx 31 \ \mathrm{yr} \left( \frac{m}{M_\odot} \right)^{1/2} \left( \frac{500 \ \mathrm{km/s}}{\sigma_T} \right) \left( \frac{75 \ \mathrm{km/s/Mpc}}{H_0} \right)^{1/2},$$

where $m$ is the mass of the microlensing object, and $\sigma_T$ is the velocity dispersion involved, including transverse velocities of both quasar and observer [Eq. (9) of Gott (1981)]. Thus, in order to explain a time scale of two weeks, one would have to invoke microlensing objects of planetary masses $m \sim 10^{-6} M_\odot$ in the extended halo of the lensing galaxy.

A somewhat weaker argument against microlensing is provided by the shape of the observed event. In the single-lens case, one would expect a very smooth and symmetric, roughly bell-shaped light curve (Paczyński 1991, Fig. 2), monotonously increasing on one side, and decreasing on the other side of the maximum. This is clearly at odds with the observed shape.

Both of the arguments about the duration and shape of the microlensing event crucially depend on the assumption of low optical depth. Since image A is approximately 6″ away from the center of the lensing galaxy, it is presumably located in a region where single-lens events dominate over double- and multiple-star microlensing. If, however, the optical depth to microlensing is moderate or nearly critical, multiple-lens events can produce arbitrary shapes in the light curves, which have not been well studied in the low-amplitude regime. At optical depth approaching unity, a microlensing event is usually caused by caustic crossing, and its duration is not directly related to the mass of the lensing object, but is limited by the projected transverse velocity of the caustic and the size of the quasar continuum region.

Clearly, the only secure way to resolve the dilemma is with observations of the image B light curve in 1996. In principle, this would still leave the possibility of accidental repeated microlens-induced features in the two images, but this is extremely unlikely in the case of the double quasar 0957+561 (Falco, Wambsganss & Schneider 1991).

## 5. Discussion

It has been widely appreciated (see, e.g., Blandford & Narayan 1992) that measurement of gravitational lens differential time delays may provide a simple (*i.e.*, understood entirely from first principles), direct and robust way of measuring absolute physical distances, and thus $H_0$, at large redshifts. It is daunting that a decade and a half of observational work on gravitational lens systems has provided so little progress. The difficulties are, in a sense, subtle; the individual astronomical observations (image photometry) are reasonably straightforward as is the theoretical interpretation of the resulting light curves.

One major problem is logistical. Obtaining small amounts of data (observing time) frequently and regularly with a large (non-solar) telescope and using the same instrumentation is inconsistent with the operational modes of conventional nighttime observatories. Aside from its direct interest, we believe that the data set presented here demonstrates that remote observing and fast instrument change technologies now make such consistent synoptic projects practical. We intend to continue to apply this methodology to 0957+561A,B and to other gravitational lens systems, some of which are likely to be more suitable for a determination of $H_0$.

In summary, we report a sharp event in the leading (A) image of the gravitational lens system 0957+561. The chromaticity, amplitude and shape of this event argue for its intrinsic nature and against microlensing-induced variability. However, microlensing cannot be reliably excluded if optical depth is close to critical. A detection of the observed feature in the light curve of the trailing (B) image in 1996 would confirm the quasar variability hypothesis and settle a long-standing controversy about the value of the gravitational time delay in the double quasar 0957+561. In addition, it would further constrain the lensing models of this system by providing the brightness ratio of the two images at optical wavelengths, and thus bring us one step closer towards the measurement of the Hubble's constant.

Hans Witt and Ralph Wijers provided valuable advice and assistance in the early stages of the project. We thank Bohdan Paczyński and Shude Mao for critical reading of the manuscript and useful comments. This demanding monitoring project would not have been possible without the expert help of observing specialists at APO, Louis E. Bergeron, Karen A. Gloria and Daniel C. Long, which often went beyond the call of duty. WNC would like to thank the Fannie and John Hertz Foundation for its continued support. This work was supported by NSF grant number AST94-19400.

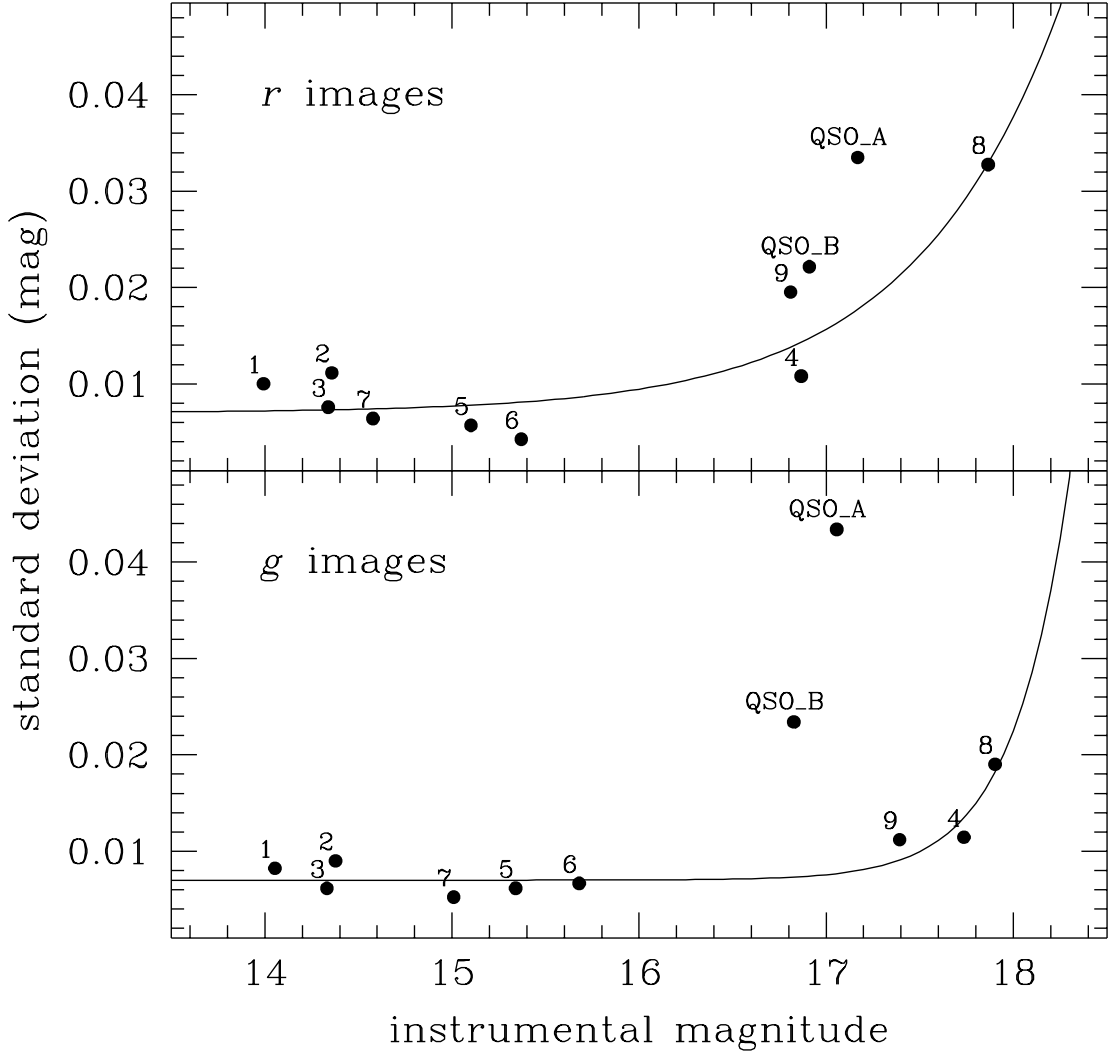

Fig. 1.— Standard deviation of instrumental magnitude $\sigma[m0(s)]$ versus instrumental magnitude $m0(s)$ for quasar images and comparison stars in $g$ and $r$ frames of the 0957+561 field. Stars are numbered 1 through 9, and quasar images are labeled QSO_A and QSO_B, where QSO_A is the northern image. A constant+exponential fit to the stars is plotted with a solid line.

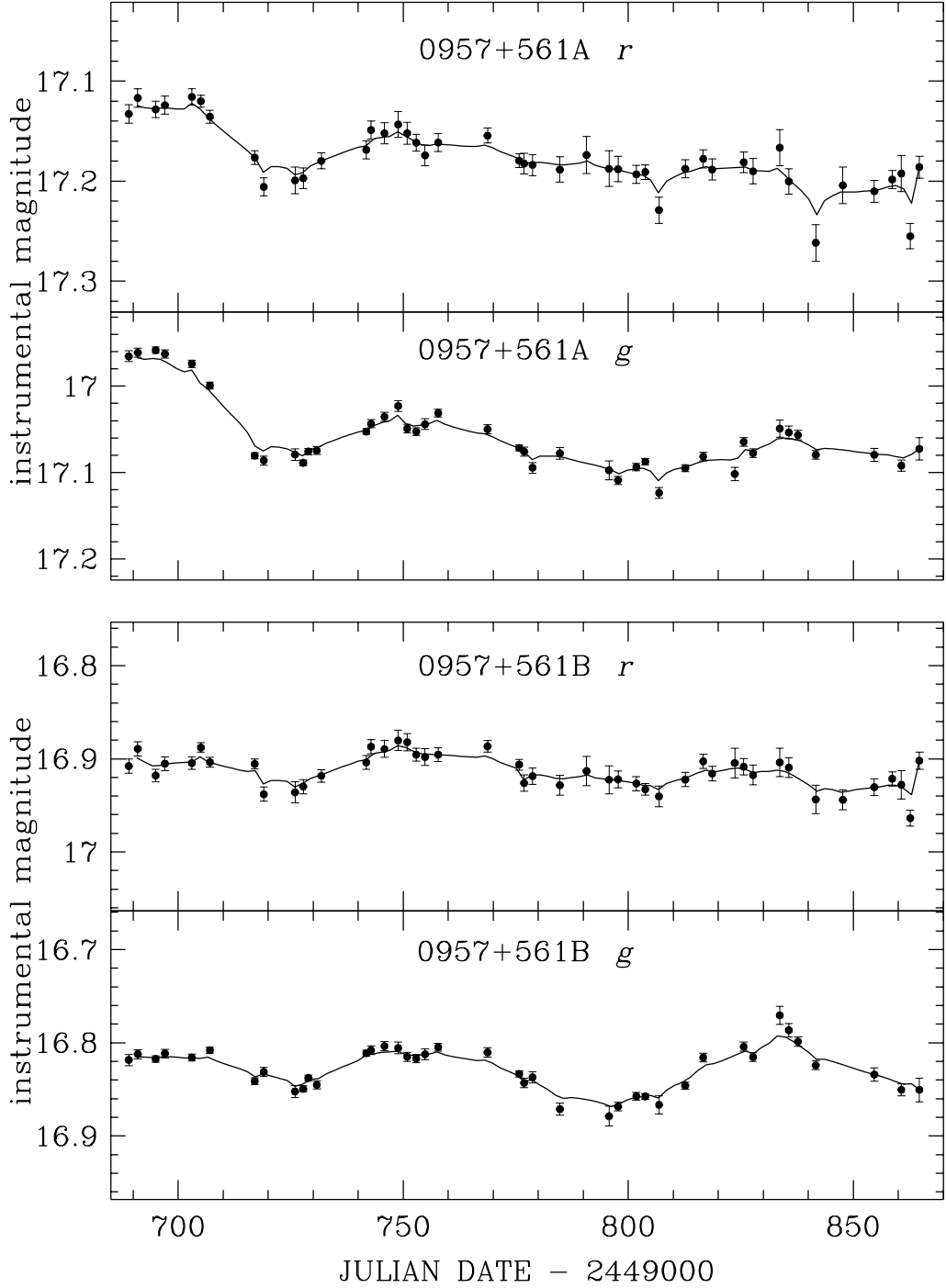

Fig. 2.— Light curves for A and B images of the QSO 0957+561 in *g* and *r* bands. Data encompasses 6 months of observations between December 2, 1994 and May 31, 1995. Multiple exposures taken on the same night were combined to a single point. A sharp drop was detected in the leading image A at the beginning of the observing season, with an approximate amplitude of 0.08 mag in *r* and 0.13 mag in *g*. The duration of the event from peak to minimum was approximately two weeks.

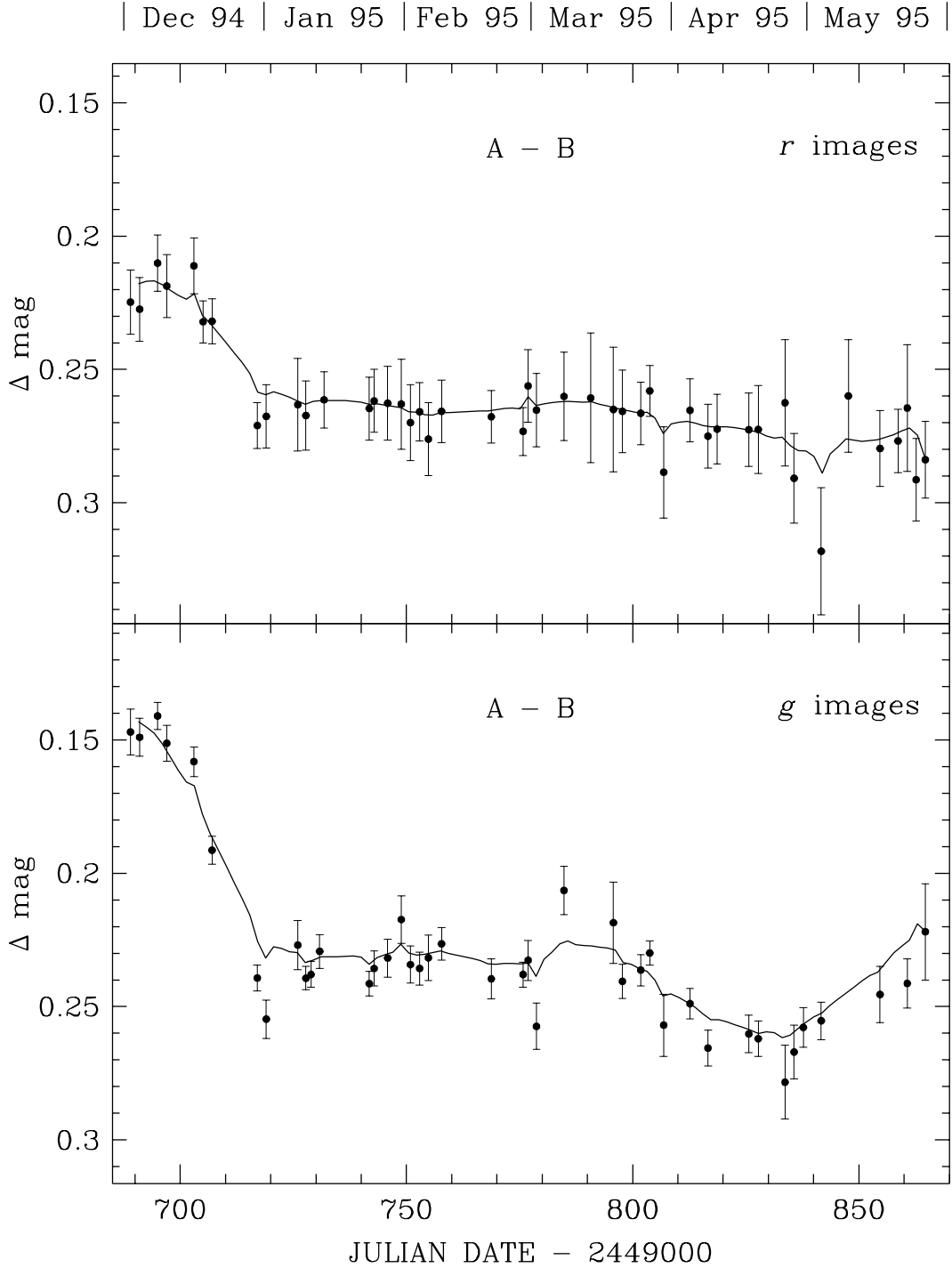

Fig. 3.— Difference in instrumental magnitudes of A and B images of QSO 0957+561 in $g$ and $r$ bands. Errorbars are taken from Fig. 2 and added in quadrature. This plot clearly shows that the event at the beginning of the observing season is not a systematic effect of correlated variations in both images, but a real change in the measured flux of image A.

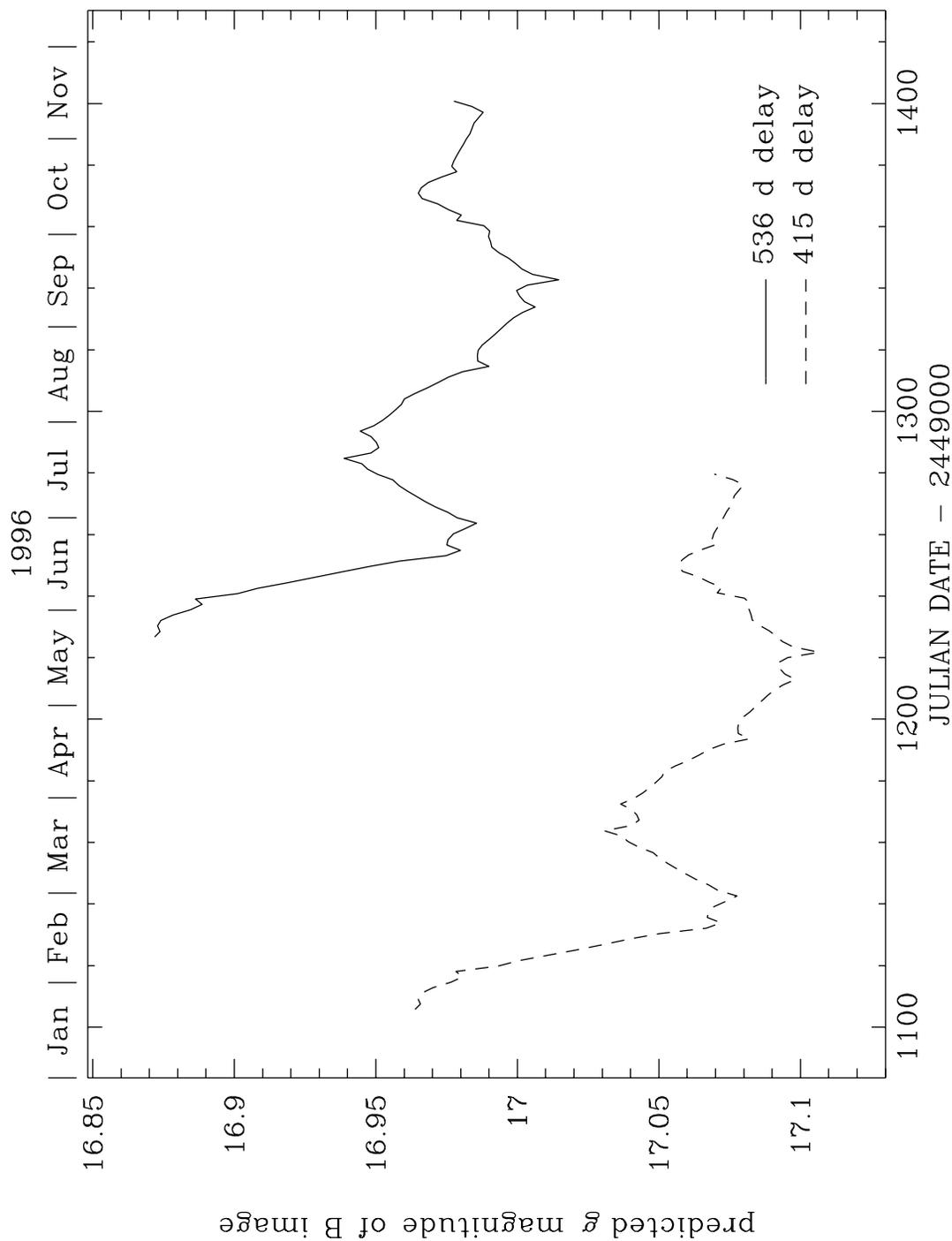

Fig. 4.— Predicted light curves for the B image of the QSO 0957+561 assuming the gravitational time delays and intrinsic image magnitude offsets reported in the literature. Solid line is the Press et al. (1992b) estimate of $\tau = 536$ days and $A - B = 0.0950$ mag, and dashed line is the Vanderriest et al. (1989) estimate of $\tau = 415$ days and $A - B = 0.03$ mag.